\documentclass[10pt,letterpaper]{article}
\pdfoutput=1
\usepackage{opex3}
\usepackage{units}
\usepackage{color}

\usepackage{textcomp}

\begin{document}

\title{Squeezed light at 1550\,nm with a quantum noise reduction of 12.3\,dB}

\author{Moritz Mehmet$^{1,2}$, Stefan Ast$^{1}$, Tobias Eberle$^{1}$, Sebastian Steinlechner$^{1}$, Henning Vahlbruch$^{1}$, and Roman Schnabel$^{1,*}$}

\address{$^1$Max-Planck-Institut f\"ur Gravitationsphysik (Albert-Einstein-Institut) and\\ Institut f\"ur Gravitationsphysik der Leibniz Universit\"at Hannover,\\ Callinstr. 38, 30167 Hannover, Germany\\
$^2$Centre for Quantum Engineering and Space-Time Research - QUEST,\\Leibniz
Universit\"at Hannover, Welfengarten 1, 30167 Hannover, Germany}

\email{moritz.mehmet@aei.mpg.de} 

\begin{abstract}
Continuous-wave squeezed states of light at the wavelength of 1550\,nm have recently been demonstrated, but so far the obtained factors of noise suppression still lag behind today's best squeezing values demonstrated at 1064\,nm. Here we report on the realization of a half-monolithic nonlinear resonator based on periodically-poled potassium titanyl phosphate which enabled the direct detection of up to 12.3\,dB of squeezing at 5\,MHz. Squeezing was observed down to a frequency of 2\,kHz which is well within the detection band of gravitational wave interferometers. 
Our results suggest that a long-term stable 1550\,nm squeezed light source can be realized
with strong squeezing covering the entire detection band of a 3rd generation gravitational-wave detector such as the Einstein Telescope.    
\end{abstract}



\section{Introduction}
It has been shown by Caves~\cite{Caves81} that squeezed light can be used to enhance the sensitivity of (quantum noise limited) interferometers as they pursue the first direct detection of gravitational waves (GWs). 
Caves' analysis revealed that the quantum noise contribution to the interferometer output signal was almost completely caused by the vacuum noise coupling into the interferometer through the open port of its 50/50 beamsplitter. The replacement of this vacuum field by a field with less quantum noise in the measured quadrature, namely a squeezed vacuum state, would directly decrease the photon shot noise on the detector and hence increase the output signal to noise ratio. 

All current GW interferometers are operated with lasers at 1064\,nm. Accordingly, squeezed light research within the GW community has mainly been focused on the generation of continuous-wave (cw) squeezed light at that respective wavelength. Maximum squeezing values of 11.5\,dB from a bulk lithium niobate (LiNbO$_3$) crystal~\cite{Mehmet2010} and 12.7\,dB from a periodically-poled potassium titanyl phosphate (PPKTP) crystal~\cite{Eberle2010} have been reported. These results were obtained in the MHz regime, and more importantly they used \emph{monolithic} resonators to reduce intra-cavity optical loss. Monolithic resonators are not feasible as a GW detector squeezed light source, because fast tuning of their resonance frequency to follow the detector's baseline laser is not possible. 
Active length/frequency stabilization can be realized by setting up the squeezed light source as a hemilithic (half-monolithic) resonator and feeding back a control signal to the piezo-driven coupling mirror. The implementation of such a control field usually introduces excess noise (typically technical laser noise) which masks the squeezing at frequencies below a few megahertz. Vahlbruch et al.~\cite{Vahlbruch06} proposed and demonstrated what they called a {\it coherent control scheme} to solve this issue. This control scheme facilitated the detection of squeezed light ranging over frequencies from 10\,kHz down to 1\,Hz~\cite{Vahlbruch07}. It was also used to construct a squeezed light source~\cite{Vahlbruch2010} that produced up to 9\,dB of squeezing and was successfully incorporated as a squeezed light add-on into the km-scale gravitational wave detector GEO600~\cite{Grote08,GEOnature}. 

With the feasibility of squeezed light injection having been demonstrated it is most likely that this technique will be part of the major upgrades planned towards the second generation of GW detectors such as Advanced Virgo~\cite{Virgo09} and Advanced LIGO~\cite{Harry09}.    

Beyond these advanced detectors, plans for third-generation instruments already exist, e.g. those presented
in the recent European design study for the so-called Einstein Telescope (ET)~\cite{ET}. To improve the displacement sensitivity in the low-frequency range it is proposed to use cryogenically cooled interferometers with silicon test masses, which would reduce the thermal noise contribution to the detector sensitivity. Consequently, these detectors would need to be driven by 1550\,nm lasers~\cite{Rowan05,Schnabel2010}. As squeezing enhancement is also envisaged for ET, cw squeezed states at 1550\,nm covering the frequencies from 1\,Hz to 10\,kHz are required to achieve a broad-band reduction of its quantum noise. 

Squeezed states of light at a wavelength of 1550\,nm have only recently been produced in the cw regime~\cite{Mehmet09a,Mehmet09b,Eberle2011} with the highest value of 9.9\,dB reported in~\cite{Eberle2011}. 
So far, squeezing factors beyond 10\,dB were reported from monolithic cavity designs and at megahertz frequencies only.
Here, we report on the observation of 1550\,nm squeezed light from an optical parametric amplifier (OPA) that was designed as a 
hemilithic resonator and generated squeezing values exceeding 10\,dB at frequencies of as low as 18\,kHz. Squeezed vacuum states were generated using quasi phase-matched degenerate parametric down-conversion, also known as optical parametric amplification, in a crystal made of PPKTP. The crystal end face and a piezo-driven coupling mirror were used to realize the squeezing resonator in the desired hemilithic configuration.
By means of balanced homodyne detection, a non-classical noise reduction of 12.3$\pm0.2$\,dB was directly observed at a sideband frequency of 5\,MHz. Furthermore, we extended our investigation to the sub megahertz-regime which revealed a squeezing factor of more than 10\,dB from 80\,kHz down to a frequency of 18\,kHz. A maximum noise suppression of 11.4$\pm 0.5$\,dB was observed between 60\,kHz and 80\,kHz. 
Despite a considerable degradation of the squeezing level from 18\,kHz downwards some degree of noise reduction expanded well into the detection band of GW detectors with a remaining 5\,dB of squeezing measured at a frequency of 2\,kHz.

\section{Experimental setup}
\begin{figure}
\begin{center}
\includegraphics[width=0.8\textwidth]{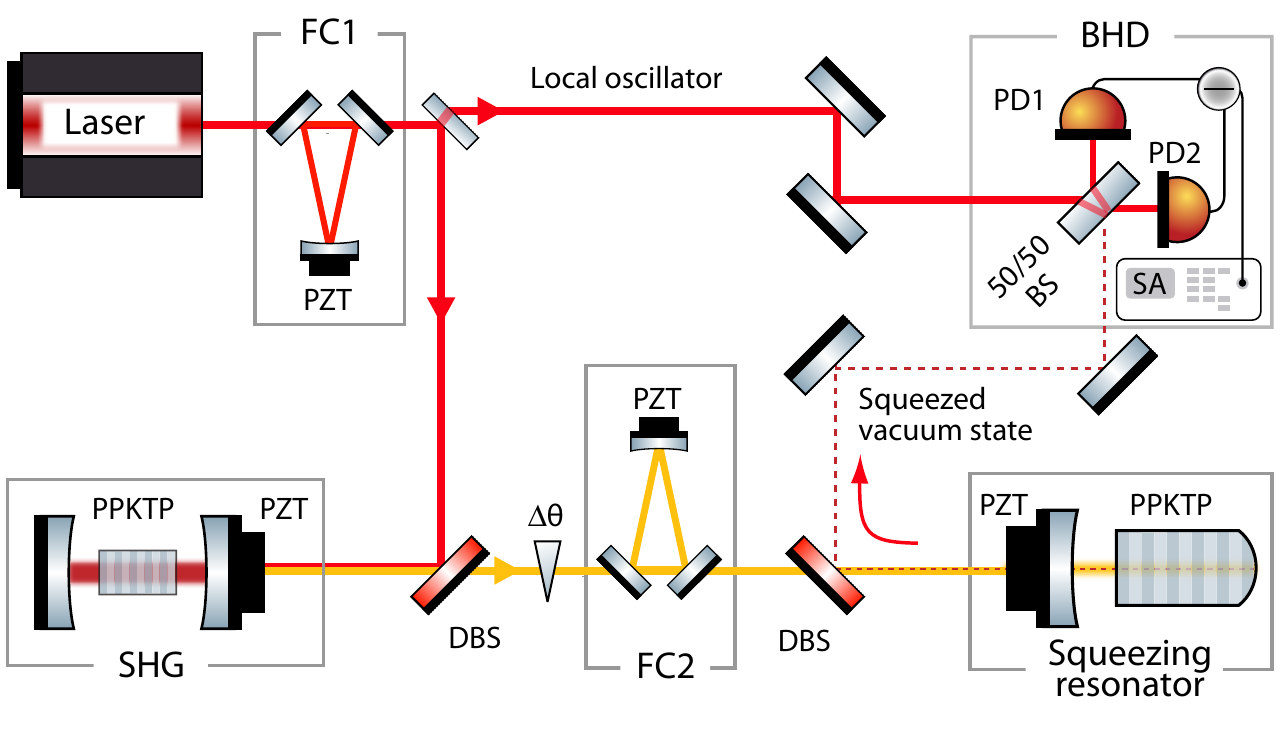}
  \caption{Schematic of the experiment. Laser: 1550\,nm fiber laser.
  FCs, filter cavities for spatio-temporal mode cleaning at both wavelengths. 
  PZT, piezoelectric transducer;
  SHG, external second-harmonic generator to produce the 775 nm pump field; BS, beam splitter;
  DBS, dichroic beam splitter; PDs, photo diodes, SA, spectrum analyzer that converts the differential current of the two photo diodes into variances; BHD, balanced homodyne detector for the characterization the squeezed field quadratures.}
 \label{Fig1}
\end{center}
\end{figure}

A schematic of our experimental setup is shown in Fig.~\ref{Fig1}.
As the laser source we used a commercially available high power cw erbium microfiber laser with an output power of 1.5\,W. The laser was transmitted through a three-mirror ring cavity (FC1) that provided spatio-temporal filtering of the beam for the downstream experiment. Approximately 20\,mW of the transmitted light were reserved to be used as the local oscillator (LO) for balanced homodyne detection (BHD). The remaining light was sent into a second-harmonic generator (SHG) to furnish the 775\,nm pump field needed to drive the OPA.%
The SHG was a single-ended standing wave cavity built by two mirrors with 25\,mm radius of curvature with a double anti-reflection (AR) coated, 10\,mm long PPKTP crystal in between. The end mirror was highly reflective (HR) for both wavelengths; the piezo-driven coupling mirror had a 90\,\% power reflectivity for the fundamental and only a marginal reflectivity ($<4$\,\%) for the second-harmonic field. Taking into account the refractive index of $n=1.816$ for PPKTP at 1550\,nm and the spacing of 20\,mm between the crystal end faces and the mirrors, the cavity waist size and linewidth were w$_{0} = 60\,\mu$m and $\textrm{FWHM} = 43$\,MHz, respectively. A second ring cavity (FC2) with specifications similar to FC1 was incorporated to reduce amplitude and phase fluctuations on the pump field,  

In order to servo-control the length of FC1, FC2 and the SHG to be resonant with their respective carrier fields a modulation-demodulation locking technique was used. The required phase modulations were imprinted onto the light using electro-optical modulators (not shown in Fig.~\ref{Fig1}) with frequencies (101.25\,MHz for FC1 and FC2, 138\,MHz for the SHG) chosen to be far outside the cavity linewidth such that unwanted sidebands on the transmitted light were kept small.
The filtered pump beam was then mode matched to the TEM$_{00}$ mode of the OPA. 

Our OPA included a piezo-actuated coupling mirror and a PPKTP crystal of dimensions 9.3\,mm$\times$2\,mm$\times$1\,mm. The crystal end face had a radius of curvature of 12\,mm and was HR coated for 1550\,nm and 775\,nm, thus serving as the cavity end mirror. The PPKTP front side was AR coated for both wavelengths. An air gap of 23\,mm length was realized between the AR coated side of the crystal and the coupling mirror. As the input/output coupler we chose a half-inch substrate that had a power reflectivity of 90\,\% for the fundamental and 20\,\% for the second harmonic field and had a radius of curvature of 25\,mm. The intra-cavity waist size and linewidth were $40\,\mu$m and 63\,MHz, respectively.
A peltier element beneath the crystal was used to keep the phase matching temperature at around 40\textdegree C.  

The squeezed states left the OPA in the opposite direction to the pump and were directed towards the balanced homodyne detector's 50/50 beam splitter upon reflection at a dichroic beam splitter (DBS).

In squeezing experiments the interference contrast between the local oscillator and the squeezed (signal) field on the 50/50 beam splitter (also known as the homodyne visibility) is crucial because any mode mismatch quadratically translates into optical loss which degrades the squeezing level. To adjust the homodyne visibility a bright beam was needed that resembled the squeezed vacuum output of the OPA. To this end an auxiliary beam (omitted in Fig.\ref{Fig1}) was matched to the TEM$_{00}$ mode through the HR rear surface of the OPA. With this control beam the OPA length could be locked and the transmitted light propagated congruent with the mode to be squeezed. By carefully adjusting the overlap between auxiliary signal beam and LO a fringe visibility of 99.5\% was achieved.
During data taking the auxiliary field was blocked. This ensured true vacuum squeezing and avoided the degradation of the squeezing by additional noise from the control beam. The rigidity of the cavity construction allowed us to keep the squeezed light source on resonance by solely applying DC-voltage to the piezo. 
\section{Experimental procedure and results}
The two observables of interest were the amplitudes of the amplitude quadrature $\hat X_1$ and the phase quadrature $\hat X_2$ of the electromagnetic field. The homodyne detector measured their respective variances which obey an uncertainty relation, $\Delta^2 \hat X_1 \cdot \Delta^2 \hat X_2 \geq \nicefrac{1}{16}$~\cite{GerryKnight}.
The relative phase $\theta$ between signal beam and local oscillator defines the quadrature under investigation, given by  $\hat X(\theta)= \hat X_{1} \cos \theta   + \hat X_{2} \sin \theta $. 
As depicted in Fig.\ref{Fig1}, in our setup a phase shifter (piezo driven mirror) in the pump path allowed us to vary the phase of the pump incident on the OPA. This led to a change of the phase with which the squeezing was produced and hence enabled us to adjust $\theta$ to measure, e.g., the squeezed quadrature ($X(0)=X_1$) or the anti-squeezed quadrature ($X(\pi/2)=X_2$), respectively. The pump power fluctuations due to beam pointing relative to the eigenmode of FC2 were below 1\% and could be neglected. The output signal of the homodyne detector, i.e., the differential current of the two photo diodes, was converted into variances with a spectrum analyzer.
\subsection{Squeezing at 5\,MHz sideband frequency}
\begin{figure}[b]
\begin{center}
\includegraphics[width=0.8\textwidth]{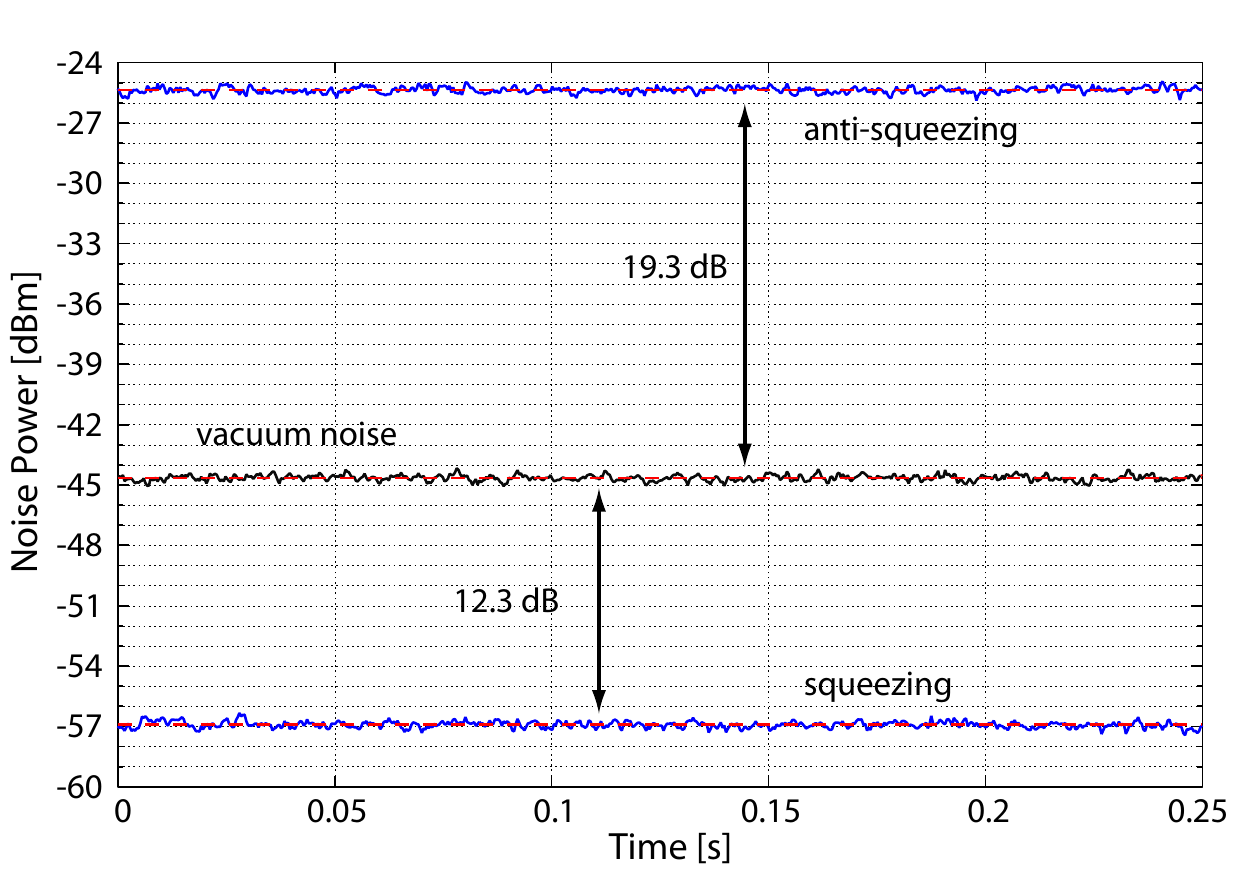}
 \caption{Balanced homodyne measurements of the quadrature noise variances. The vacuum reference was recorded with a LO power of 11.8\,mW and a blocked signal port. The OPA was driven with a pump power of approximately 180\,mW. With an open signal port, the squeezed variance and the anti-squeezed variance were measured by choosing the corresponding relative phase between signal field and LO.
The depicted measurements are averages of 2 traces that were recorded at a Fourier frequency of 5\,MHz, with a resolution bandwidth of 200 kHz and a video bandwidth of 200 Hz. No data postprocessing was applied, i.e., the data still include electronic dark noise, and thus represent direct observations. 
A linear fit to the linearized variances was used to determine the variance mean values and the associated standard deviation. The anti-squeezing was at +19.3$\pm0.2$\,dB relative to the vacuum noise, whereas the squeezing was at -12.3$\pm0.2$\,dB.}
\label{zeroSpan}
\end{center}
\end{figure}
We performed measurements at a sideband frequency of 5\,MHz and with a pump power of approximately 180\,mW. The data as collected by the spectrum analyzer are shown in Fig~.\ref{zeroSpan}. Approximately 11.8\,mW of LO power was used to obtain the vacuum (or shot noise) reference level with the signal input port of the 50/50 beam splitter blocked.  When the input port was opened, the phase $\theta$ was subsequently adjusted to measure the anti-squeezed and the squeezed quadrature. A squeezing value of 12.3$\pm0.2$\,dB below shot noise was directly measured. The corresponding anti-squeezing was 19.3$\pm0.2$\,dB above the shot noise level. The homodyne detector's electronic dark noise was 26\,dB below the vacuum noise and was not subtracted from the data. 

\subsection{Characterization of the OPA}
For the below-threshold OPA the anticipated variances of the squeezed and anti-squeezed quadratures ($V_{1}$ and $V_{2}$, respectively) can be modeled by \cite{Aoki06} 
\begin{equation}\label{sqzspec}
    V_{1,2} = 1\pm \eta \frac{4 \sqrt{P/P_{\textrm{thr}}}}{\left(1\mp \sqrt{P/P_{\textrm{thr}}}\right)^2+4 \left(  2 \pi f \kappa^{-1} \right  )^2}\ ,
\end{equation}
where $\eta$ is the total detection efficiency (in terms of optical loss) which, when non-perfect, impairs the squeezing. $P$ is the second-harmonic pump power and $P_{\textrm{thr}}$ is the amount of pump power needed to reach the OPA threshold. The cavity round-trip length $l$, the coupling mirror transmissivity $T$, and the round trip loss $L$ yield the cavity decay rate $\kappa= c(T+L)l^{-1}$; $c$ is the speed of light and $f$ is the measurement frequency. 

\begin{figure}
\begin{center}
\includegraphics[width=0.8\textwidth]{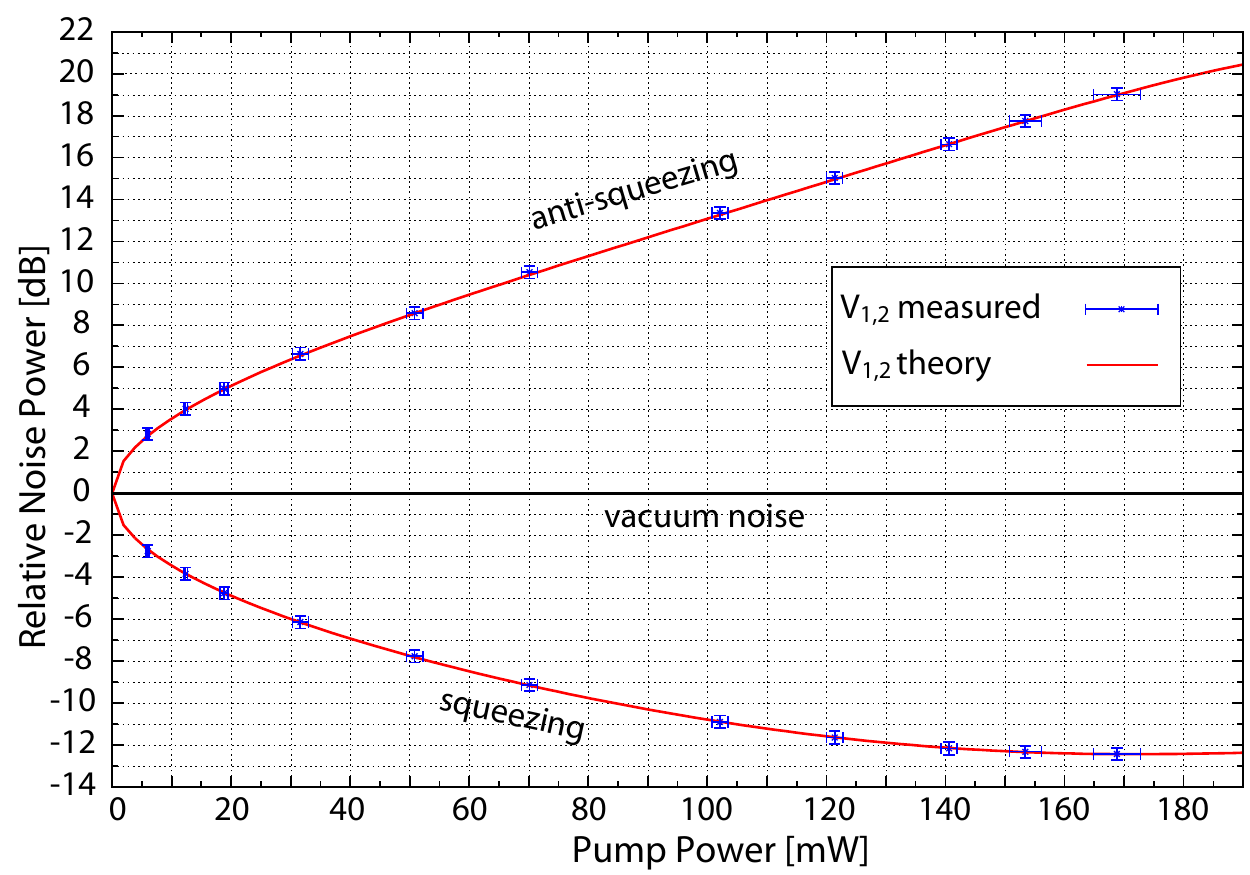}
  \caption{Pump power dependence of anti-squeezed and squeezed quadrature variances. All values were obtained from zero-span measurements at 5\,MHz. In order to fit the numerical model, all the data were dark-noise corrected and subsequently normalized to the vacuum reference.}
 \label{SQZvsPUMP}
\end{center}
\end{figure}

The amount of detectable squeezing is not limited by optical loss alone. Fluctuations in the relative phase between signal and LO can likewise degrade the measured squeezing level. In an experiment, such phase noise might have numerous origins, for example unstable locking loops, residual phase modulation, or acoustically induced mirror motion. When assuming the standard deviation of the normally distributed phase fluctuations to be small, phase jitter with an rms of $\theta_{\mathrm{fluc}}$ is equal to the homodyne detector measuring at a phase offset $\theta_{\mathrm{fluc}}$. The resulting variances can then be written as    
\begin{equation}\label{sqzspecPN}
 V^\prime_{1,2} = V_{1,2} \cos^2{\theta_{\mathrm{fluc}}} + V_{2,1} \sin^2{\theta_{\mathrm{fluc}}}\ .
\end{equation}
At a Fourier frequency of 5\,MHz, we repeatedly took zero-span measurements of squeezing and anti-squeezing at different pump powers. The collected data are shown in Fig.~\ref{SQZvsPUMP}. 
All measurements were normalized to their respective vacuum level. Here, the contribution of the electronic detector dark noise was subtracted and hence the value of best (optimum) squeezing improved to $-12.4 \pm 0.3$\,dB. The horizontal error bars correspond to the fluctuation of the pump power $P$ during data taking. For constant monitoring of $P$, a calibrated photo diode in transmission of a steering mirror in the pump path was used. The absolute error of a given pump power is $\pm3\,\%$ due to the measurement uncertainty of the power meter used for calibration. The vertical error bars are the standard deviation of 0.3\,dB as given by the linear fit to each recorded trace.
Taking into account the parameters of our cavity, namely $l=7.98$\,cm,  $T=10$\,\%, $L= 0.1$\,\%, and the measurement frequency $f=5$\,\,MHz, we fitted for the total detection efficiency, the threshold power and the amount of phase noise using our model defined in Eqs.~(\ref{sqzspec}) and~(\ref{sqzspecPN}).
This yields the following values: $\eta=0.965\pm0.002 $, $P_{\textrm{thr}}=221\pm 3$\,\,mW, $\theta_{\textrm{fluc}}=0.66\pm0.06${\textdegree}.   
\begin{figure}
\begin{center}
\includegraphics[width=0.8\textwidth]{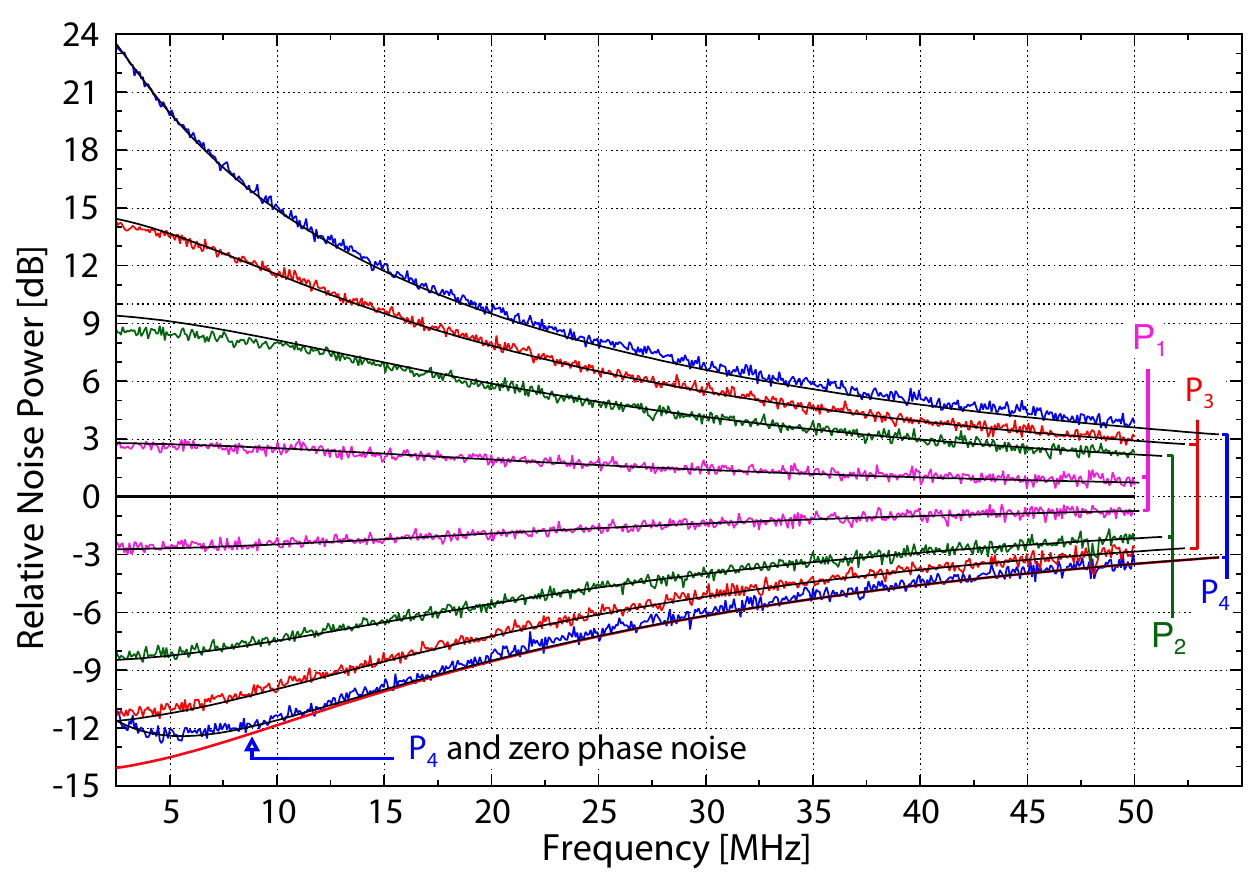}
  \caption{Pump power dependence of the squeezing spectra, experiment and theory. All depicted traces are averages of 10 individual traces each recorded with a resolution bandwidth $\textrm{RBW}=500$\,kHz and video bandwidth $\textrm{VBW}=2$\,kHz. All traces were corrected for electronic detector dark noise and were normalized to the vacuum level. The theoretical predictions (solid lines) were obtained from Eq.~(\ref{sqzspecPN}) using $\eta=0.965$, $P_{\textrm{thr}}=221$\,mW, and $\theta_{\textrm{fluc}}=0.66${\textdegree}  with the respective pump power values of P$_1=6$\,mW, P$_2=56$\,mW, P$_3=106$\,mW, and P$_4=180$\,mW. The remaining cavity parameters were used as given in the text. The bottom curve shows the squeezing spectrum under the assumption of zero phase noise.}
 \label{SqzSpec}
\end{center}
\end{figure}
We confirmed these results by inserting the obtained values into Eq.~(\ref{sqzspecPN}) to model the frequency dependence of our squeezer and compared it to measured squeezing spectra. For that we took measurements from 2.5\,MHz to 50\,MHz at several pump powers between 6\,mW and 180\,mW. The collected data are depicted in Fig.~\ref{SqzSpec}. All spectra were averaged 10 times. 
As can be seen in Fig.~\ref{SqzSpec} our theoretical model does indeed provide a good prediction of the experimental outcome. 
We especially note the quality of the fit with regards to the phase noise parameter $\theta_\textrm{fluc}$ since a considerable degradation of the squeezing
level at frequencies below 5\,MHz becomes apparent when approaching high pump powers.
Assuming zero phase noise, the squeezing level would approach 14\,dB as indicated by the bottom curve in Fig.~\ref{SqzSpec}.  
\subsection{Low frequency performance}
\begin{figure}[h]
\begin{center}
\includegraphics[width=0.8\textwidth]{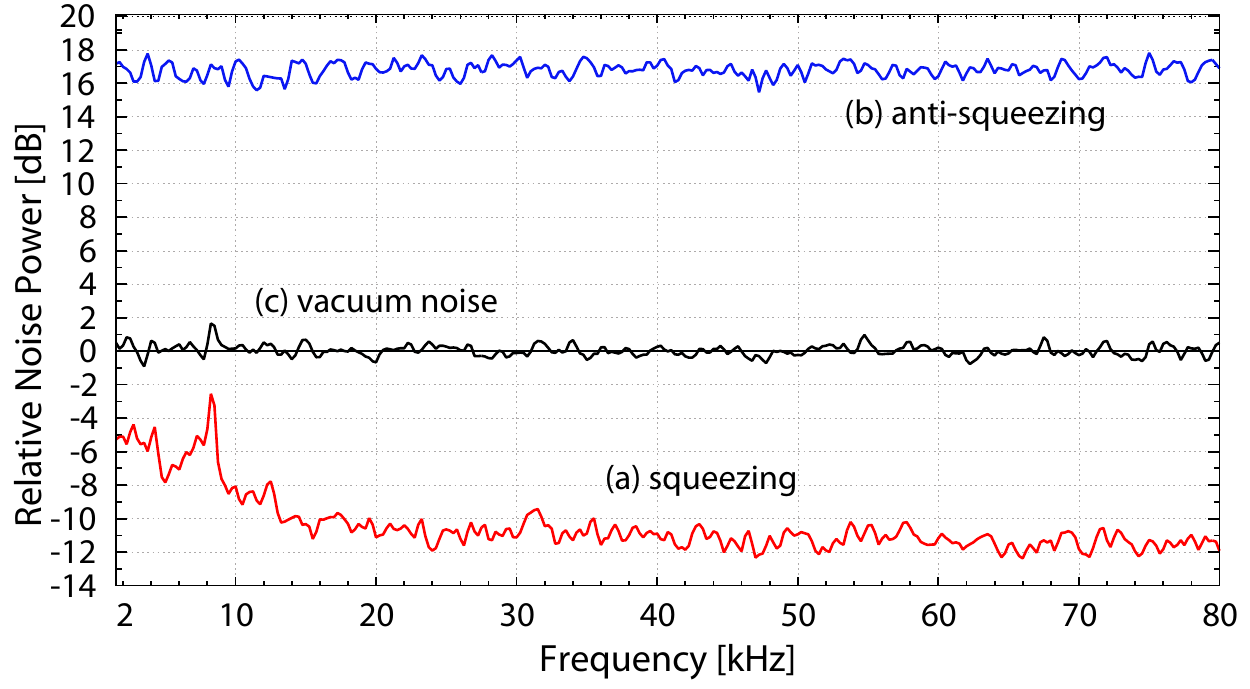}
\caption{Squeezing at frequencies between 1.5\,kHz and 80\,kHz. All traces were measured with RBW of 250\,Hz and were normalized to the vacuum noise level corresponding to a LO power of 4.8\,mW. Each measurement point is the averaged root mean square value of 20 [40 for trace(c)] measurements. Trace (a) shows the direct observation of more than 10\,dB of squeezing down to a frequency of 18\,kHz with a maximum noise suppression of $-11.4 \pm0.5$\,dB between 60\,kHz and 80\,kHz. Below 18\,kHz the squeezing level degraded, reaching approximately $-5$\,dB at 2\,kHz. The peak at 8\,kHz was caused by a residual, non-stationary amplitude modulation from FC1. The anti-squeezing [trace (b)] was 16.8$\pm0.5$\,dB above the vacuum noise level as measured by the homodyne detector [trace (c)]. Electronic dark noise was not subtracted from the measurement data.}
\label{lowFsqz}
\end{center}
\end{figure}

For the observation of squeezing in the kHz regime, we used our homodyne detector together with a low-noise FFT-analyzer. 
This enabled us to measure the quantum noise from 80\,kHz downwards. 
For all measurements a power of 4.8\,mW for the LO was used. This yielded a vacuum reference level with a clearance from the electronic dark noise of approximately 15 to 20\,dB over the entire FFT window. Therefore, no data postprocessing to subtract the detector dark noise was applied. The resulting noise spectra are shown in Fig.~\ref{lowFsqz}.   
The measured anti-squeezing [trace~(b)] was at 16.8$\pm0.5$\,dB above the vacuum noise level  [trace~(c)]. The spectrum of the squeezed quadrature [trace~(a)] showed a maximum non-classical noise suppression of $-11.4\pm0.5$\,dB between 60\,kHz and 80\,kHz. Going down in frequency the squeezing slightly decreased reaching $-10.7\pm0.4$\,dB at 20\,kHz. From 18\,kHz downwards the squeezing level degraded considerably. We attribute this mainly to two characteristics of our setup. First, no stabilization scheme was used to control the relative phase between LO and signal field. Hence, acoustically excited motion cannot be compensated for and together with the prolonged measurement times at smaller sideband frequencies this led to an averaging effect that decreased the observed squeezing level. Second, no Faraday isolator was incorporated between the squeezing resonator and homodyne detector to protect the OPA from back scattered/reflected light. It is known that stray light can introduce excess noise~\cite{Bowen02,McKenzie04} and can cause parasitic interferences~\cite{Vahlbruch07} within the system. Both effects can impair or even completely annihilate squeezing in the audio band. Despite the roll-up towards lower frequencies, squeezing of approximately $-5$\,dB could still be observed at 2\,kHz.
\section{Conclusion}
By means of balanced homodyne detection we directly observed 12.3\,dB of squeezing of the vacuum fluctuations of a 1550\,nm cw laser beam at a sideband frequency of 5\,MHz. This non-classical noise reduction was achieved by degenerate parametric down-conversion below threshold in a standing wave resonator. High quality optics and the low absorption of the PPKTP crystal made it possible to construct the squeezing resonator as a hemilithic cavity and still keep the optical loss small enough to detect such strong squeezing. Pump power dependent squeezing values were used to characterize our squeezer in terms of optical loss and phase noise.  
By fitting a theoretical model to the data the overall optical loss and residual phase noise between signal and LO were determined to be 3.5\% and 0.66{\textdegree}, respectively. Note that the value obtained for the optical loss includes the escape efficiency of the squeezing resonator, propagation loss and the subsequent homodyne detection.  
The amount of phase noise could most likely be reduced by the implementation of a stable control for the phase of the pump beam as well as the homodyne detection phase. Assuming zero phase noise, 14\,dB of (dark noise corrected) squeezing would then be feasible in our setup. However, such control schemes are usually not 
easily realized without introducing excess noise from the participating light fields and hence were not pursued here.  

Squeezed vacuum states at frequencies within the detection band of GW interferometers can be used to enhance the sensitivity of the respective detector. 
These frequencies typically range from some Hertz up to 10\,kHz. Here, we were able to measure up to 11.4\,dB of squeezing below 80\,kHz.
Even though from 18\,kHz downwards acoustically induced modulation and residual stray light degraded the level of squeezing, it extended into the detection band of a GW detector. Owing to the hemilithic layout of the squeezed light resonator, a coherent control scheme according to~\cite{Vahlbruch06} is applicable. With such a control scheme the squeezing spectrum would presumably broaden to cover the entire range of frequencies down to a few hertz. Additionally, the required long-term stability would be obtained and an even higher squeezing factor can be expected since the aforementioned phase noise would be reduced.

The authors acknowledge the IMPRS on Gravitational Wave Astronomy for support.
\end{document}